# Shubnikov-de Haas-like Quantum Oscillations in Artificial One-Dimensional LaAlO$_3$/SrTiO$_3$ Electron Channels

Guanglei Cheng[1,2,3,5], Anil Annadi[2,3], Shicheng Lu[2,3], Hyungwoo Lee[4], Jung-Woo Lee[4], Mengchen Huang[2,3], Chang-Beom Eom[4], Patrick Irvin[2,3] and Jeremy Levy[2,3*]

[1]CAS Key Laboratory of Microscale Magnetic Resonance and Department of Modern Physics, University of Science and Technology of China, Hefei 230026, China

[2]Department of Physics and Astronomy, University of Pittsburgh, Pittsburgh, PA 15260, USA.

[3]Pittsburgh Quantum Institute, Pittsburgh, PA, 15260, USA.

[4]Department of Materials Science and Engineering, University of Wisconsin-Madison, Madison, WI 53706, USA.

[5]Hefei National Laboratory for Physical Sciences, Hefei 230026, China

[*] E-mail: jlevy@pitt.edu

**Abstract**:

The widely reported magnetoresistance oscillations in LaAlO$_3$/SrTiO$_3$ heterostructures have invariably been attributed to the Shubnikov-de Haas (SdH) effect, despite a pronounced inconsistency with low-field Hall resistance measurements. Here we report SdH-like resistance oscillations in quasi-1D electron waveguides created at the LaAlO$_3$/SrTiO$_3$ interface by conductive atomic force microscopy lithography. These oscillations can be directly attributed to magnetic depopulation of magnetoelectric subbands. Our results suggest that the SdH oscillations in 2D SrTiO$_3$-based systems may originate from naturally forming quasi-1D channels.

SrTiO$_3$-based interfaces, and in particular the LaAlO$_3$/SrTiO$_3$ (LAO/STO) interface [1], combine the motif of semiconductor heterostructures such as GaAs/AlGaAs, with the wide-ranging physical phenomena of complex-oxides. The LAO/STO system exhibits a wide range of gate-tunable phenomena including superconductivity [2,3], magnetism [4], spin-orbit coupling [5,6] and electron pairing without superconductivity [7]. Transport at LAO/STO interfaces is complicated by a ferroelastic transition within STO at $T$=105 K [8,9]. A variety of local probe methods have documented how ferroelastic domains strongly break translational invariance and favor highly anisotropic transport along ferroelastic domain boundaries [10-13].

Given the relatively high carrier densities of oxide heterostructures, prospects for achieving quantum Hall phases under reasonable laboratory conditions appear remote. The electron density of STO-based heterostructures is approximately two orders of magnitude higher ($n_{2D}$~$10^{13}$-$10^{14}$ cm$^{-2}$) than typical III-V heterostructures, while the electron mobility is significantly lower ($\mu$~$10^3 - 10^4$ cm$^2$/Vs). A typical carrier density at the 2D LAO/STO interface is $n_{2D} = 5 \times 10^{13}$ cm$^{-2}$, nearly half electron every nm$^2$, with 100 filled Landau levels are expected for a typical laboratory-achievable magnetic field $B$=10 T.

Despite the low mobility, STO-based heterostructures show distinctive features of quantum transport. The most striking signature are quantum oscillations which have been widely attributed to the Shubnikov-de Haas (SdH) effect, a precursor to the integer quantum Hall effect. The frequency and temperature dependence of SdH oscillations reveal critical information about the electron density, carrier mobility, quantum scattering time and effective mass [14]. For the STO based systems, SdH oscillations have been extensively reported [15-28]. However, major discrepancies have been observed among all the reports in the literature, as summarized below.

i) The carrier density extracted from SdH oscillations ($n_{2D}^{SdH} \sim 10^{12} cm^{-2}$) is significantly lower than what is determined by low-field Hall measurements ($n_{2D}^{Hall} \sim 10^{13} - 10^{14}$ cm$^{-2}$). Values of $n_{2D}^{Hall}/n_{2D}^{SdH}$ can range between slightly over 1 [29] to 25 [18] for $\delta$-doped STO.

ii) The number of distinct SdH frequencies for a given device is found to vary from one [15,17,19] to four [24], with considerable variation in the effective mass of the high-mobility carriers. The corresponding band assignment also ranges from $d_{xy}$ orbital [21] to $d_{xz}/d_{yz}$ orbital [27] to hybridization of the two orbitals [24].

In short, the overall phenomenology of SdH oscillations is widely reported; however, there is a lack of internal consistency among quantitative values for $n_{2D}^{Hall}/n_{2D}^{SdH}$ or agreed explanation for the discrepancies. Disagreements between $n_{2D}^{Hall}$ and $n_{2D}^{SdH}$ of more than a few percent in III-V ssemiconductor heterostructures are unusual. The large and variable deviations in expected values for $n_{2D}^{Hall}/n_{2D}^{SdH}$ have not been satisfactorily explained.

Here we report SdH-like quantum oscillations in an artificial quasi-1D electron channel formed at the LAO/STO interface. As a function of magnetic field and electrical gating, we observe a transition between quasi-1D to quasi-2D behavior, and extract characteristic parameters such as the carrier density and width. These electron waveguides exhibit clean ballistic transport at carrier densities as low as 320 μm$^{-1}$ (or equivalently $8 \times 10^{11}$ cm$^{-2}$ by considering 40 nm characteristic width), significantly lower than what has been reported for bulk 2D heterostructures. The magnetotransport shows characteristic SdH-like oscillations that arise due to magnetic depopulation of the magneto-electric subbands within the quasi-1D channel. All the carriers can be fully accounted for due to the observation of conductance quantization, unlike the 2D case where carrier-density measurements of SdH oscillations and Hall effect disagree. We compare the well-understood electron transport within these artificial electron waveguides with naturally formed channels that have been found to arise at ferroelastic domain boundaries, and suggest that previously reported SdH oscillations are in fact manifestations of naturally formed quasi-1D channels.

The electron waveguides are fabricated using conductive atomic force microscopy (c-AFM) lithography at the LAO/STO interface, as described in Ref. [30]. Starting from an insulating 3.4-unit-cell-LAO/STO interface, nanoscale conducting paths can be "written" through a surface protonation process [31,32]. Tunnel barriers or insulating regions are created through an "erasure" procedure involving surface deprotonation. Novel properties such as electron pairing without superconductivity [7] and tunable electron-electron interactions [33] have previously been revealed by studying quantum transport in similarly designed nanostructures.

Figure 1(a) shows a schematic of the overall device structure. The electron waveguide consists of a 50-nm long linear segment surrounded by two narrow insulating barriers, coupled to source and drain electrodes. The electron density within the waveguide depends on the side-gate voltage $V_{sg}$ and the overall electrostatic confinement produced by the c-AFM writing process. In the regime considered here, the cyclotron frequency $\omega_c = e|B|/m_e$ is greater than the characteristic frequency of lateral confinement $\omega_y$ ($\omega_c > \omega_y$) in magnetic fields ($B$>~3 T), where $e$ and $m_e$ are electron charge and effective mass.

Well-resolved magnetoresistance oscillations are observed at higher magnetic fields over a range of gate voltages $V_{sg}$~$0 - 100$ mV; one such example appears in Fig. 1(b). Following the analysis of previous reports [15-24,26-28], we subtract a smooth background to reveal resistance oscillations that are clearly visible and periodic in $1/B$ [Fig. 1(c)]. A Fourier analysis shows a sharp peak 26 T with a smaller secondary peak at 7 T.

The magnetoresistance oscillations clearly resemble the SdH effect; however, for an electron waveguide, it is more appropriate to consider the total *conductance*, which can be subject to Landauer quantization in the ballistic regime. Within this framework, the conductance is given by $G = (e^2/h) \sum_i T_i(\mu)$, whereby each energy subband that is occupied (at a given chemical potential $\mu$) contributes $e^2/h$ (with transmission probability $T_i(\mu)$) to the overall conductance $G$ [34]. Within this framework, the conductance increases in steps of $e^2/h$ every time the chemical potential crosses a subband energy minimum. Magnetic fields can depopulate these subbands, leading to an overall decrease in conductance with increasing field strength. Figure 2(a) shows the same data as Fig. 1, plotted as conductance in units of $e^2/h$, as a function of $V_{sg}$. Clear conductance steps are observed, while the overall slope decreases with increasing magnetic field. The conductance versus magnetic field at fixed gate voltage [Fig. 2(b)] shows quantization at integer values of $e^2/h$, with "oscillations" occurring as magneto-electric subbands are depopulated with increasing field strength. The energy spacing $\hbar\omega$ in a magnetic field, where $\omega = \sqrt{\omega_c^2 + \omega_y^2}$, is dominated by Landau level spacing $\hbar\omega_c$ in the regime where $\omega_c \gg \omega_y$. With increasing magnetic field, the occupation of the subbands is gradually depopulated as $\hbar\omega$ grows [Fig. 2(b)]. For example, at $V_{sg} = 60$ mV, the number of occupied subbands is reduced from 10 to 3 by increasing $B$ field from 0 T to 9 T. Finite-bias spectroscopy ($V_{sd}$) [Fig. 2(c)] at $B$=9 T shows clustering of *I-V* curves and half-plateaus, which provide yet another confirmation of ballistic electron transport.

In 2D electron systems, Landau levels form flat bands; the condition to completely fill level $n$ satisfies $n = \frac{\pi \hbar}{|e|B} n_{2d}$, which yields $n \propto 1/B$. In a quasi-1D electron waveguide, however, lateral confinement causes the magneto-electric subbands to exhibit a parabolic shape. The filling of the $nth$ subband is given by [35]

$$n \approx c \cdot \left(\frac{n_{2d}}{l_y}\right)^{\frac{2}{3}} \omega_y^{\frac{2}{3}}/\omega, \qquad (1)$$

where $c = \left(\frac{3}{4}\pi\right)^{\frac{3}{2}} \left(\frac{\hbar}{2m_e}\right)^{\frac{1}{3}}$ is a constant and $l_y$ is the characteristic width of the electron waveguide which determines the lateral confinement $\omega_y = \frac{\hbar}{m_e l_y^2}$. From Eq. (1), it is clear that the dependence of $n$ on $1/B$ is nonlinear in low magnetic fields and crosses over to a linear regime as $\omega \to \omega_c$ in high magnetic fields. By fitting Eq. (1), it is possible to extract key electron waveguide parameters, including the effective 2D carrier density $n_{2d}$ and width $l_y$. At higher densities, population of higher-energy vertical waveguide modes is expected, leading to new series of SdH oscillations, one for each vertical subband.

The subband index $n_i$ is assigned by referencing the resistance minima after subtracting a smooth background in varying magnetic field and $V_{sg}$, as shown in Fig. 3. Generally, the waveguide width increases with increasing $V_{sg}$. The lateral confinement frequency $\omega_y$ is large for small $V_{sg}$, leading to a saturation in the number (~ 3 in $V_{sg} = 10$ mV) of depopulated subbands in 0~9 T $B$ field range, compared to the high $V_{sg}$ case (~7 in $V_{sg} = 60$ mV). Figure 3(b) shows the dependence of $n_i$ on $1/B$. Indeed, the relationship is linear at higher magnetic fields and highly non-linear at low magnetic fields, suggesting a crossover between 2D and 1D in the electron waveguide when tuning the magnetic field.

Fitting to Eq. (1) shows good agreement with data [Fig. 3(b)]. The effective 2D carrier density increases from $n_{2D} = 8 \times 10^{11}$ cm$^{-2}$ to $n_{2D} = 2.4 \times 10^{12}$ cm$^{-2}$ as the gate voltage is increased from $V_{sg}$=20 mV to $V_{sg}$=100 mV (Fig. 4). It is worth noting that the electron density is one order of magnitude lower than what is typically reported for 2D interfaces, and two orders of magnitude lower than what is predicted for the "polar catastrophe" model.

More insights into the electronic properties of the interface can be obtained from the transconductance map d$G$/d$V_{sd}$ [Fig. 3(c)] which shows the subband structure. The lateral confinement energies $E_y = \hbar\omega_y \sim 100\ \mu eV$ and $E_z = \hbar\omega_z \sim 500\ \mu eV$ can be readily read out. Since the confinement frequency $\omega_{y,z} = \frac{\hbar}{m_{y,z} l_{y,z}^2}$, the in-plane effective mass $m_y = 0.5\ m_e$ and out-of-plane effective mass $m_z = 2.4\ m_e$ can be extracted by taking $l_y = 40\ nm$ at $V_{sg} = 20\ mV$ and a typical $l_z = 8\ nm$ [30]. Other parameters including the pairing strength and $g$-factor can be also routinely extracted [30].

The phenomena of SdH-like oscillations observed in artificially constructed quasi-1D waveguides bears a remarkable resemblance to many reports of SdH oscillations at the 2D LAO/STO interface. We propose that SdH oscillations reported for the 2D LAO/STO interface can also be accounted for by 1D transport. The ferroelastic domain patterns that have been revealed by various imaging techniques including scanning SQUID microscopy [11], scanning SET [10] and low-temperature SEM [12] points to a highly inhomogeneous landscape that is markedly different from analogous III-V semiconductor heterostructures. It is quite plausible that the transport along these ferroelastic domain boundaries may be significantly more ballistic than previously assumed. In that case, naturally forming ferroelastic domain walls would spontaneously lead to transport behavior that may vary from one device to another, and from one cooldown to another. Furthermore, inhomogeneous broadening of SdH oscillations would be expected if there are spatial variations in lateral confinements.

Inconsistencies in the number of distinct SdH oscillation frequencies among different reports can be understood by recognizing that electron waveguides can support both vertical and lateral modes. Analysis of a waveguide under harmonic lateral and vertical confinement [30] reveals a family of modes, each of which is expected to yield SdH-like quantum oscillations, with one distinct frequency for each vertical mode. For example, Fig. 1(d) shows two distinct frequencies.

The discrepancy in carrier density measurements between SdH oscillations and Hall measurements has been attributed to a number of factors at the 2D LAO/STO interface. For example, it has been suggested that coexisting high- and low-mobility carriers occupy different Ti $d$ orbitals or multiple subbands [27]. However, SdH oscillations with a single frequency have been reported below the critical density ($1.6 \times 10^{13} \mathrm{cm}^{-2}$) of Lifshitz transition [36], where only $d_{\mathrm{xy}}$ orbital is supposed to be occupied [17]; in that work, $n_{\mathrm{2D}}^{\mathrm{Hall}}/n_{\mathrm{2D}}^{\mathrm{SdH}}$ ~5. In the work reported here, the carrier density extracted by magnetic depopulation is consistent with the SdH oscillation measurement in the literature ($n_{\mathrm{2D}}^{\mathrm{SdH}} \sim 10^{12} \, cm^{-2}$). Furthermore, all of the carriers are accounted for since full conductance quantization is observed, i.e., there is no discrepancy in the carrier density measurements.

To extend the applicability of these results in a single nanowire to the more complex geometry in most 2D experiments, one needs to consider how an ensemble of ferroelastic domains would influence 2D transport. Overall, there is growing evidence that 1D channels have anomalously high mobility [37] and are ballistic on micrometer scales [38]. Recent work by

Frenkel et al. [13] indicate that anisotropic flow in 100 μm-scale device can be as large as 50% at low temperature. Devices such as these are expected to exhibit quasi-ballistic transport along ferroelastic domain boundaries, and therefore will be subject to resistance oscillations associated with magnetic depopulation of 1D subbands. Future experiments that combine one or more spatially resolved measurements with high magnetic fields could help to spatially resolve the regions that are contributing maximally to resistance oscillations in 2D structures.

In summary, we have used conductive AFM lithography at the LAO/STO interface to create quasi-1D channels whose magnetotransport characteristics bear a strong resemblance to 2D SdH-like transport widely reported at the LAO/STO interface. By analyzing the results within the framework of quantum channels, inconsistencies in accounting for "missing electrons" are resolved. Our results suggest that the non-uniform distribution of carriers due to naturally formed domain walls at the 2D LAO/STO interfaces brings additional confinements of carriers, which cause discrepancies in analyzing SdH oscillations at the 2D interfaces. More experiments are clearly required to investigate whether this framework can fully explain quantum oscillations observed at the LAO/STO interface, and to characterize mesoscopic structures in which transport along naturally formed ferroelastic domains exist. But so far, this framework has been the only one that can account for all of the observed phenomena.

**Acknowledgements:**

We thank David Pekker, Anthony Tylan-Tyler and Yun-Yi Pai for helpful discussions. This work was supported by ONR N00014-15-1-2847 (J.L.), AFOSR FA9550-12-1-0057 (J.L., C.B.E.), and grants from the National Science Foundation DMR-1104191 (J.L.), DMR-1124131 (C.B.E., J.L.). The work at University of Wisconsin-Madison was supported by the US Department of Energy (DOE), Office of Science, Office of Basic Energy Sciences (BES), under award number DE-FG02-06ER46327. The work at the University of Science and Technology of China is supported by Chinese 1000 Talents Plan for Young Scholars and the Fundamental Research Funds for the Central Universities.

**Figures:**

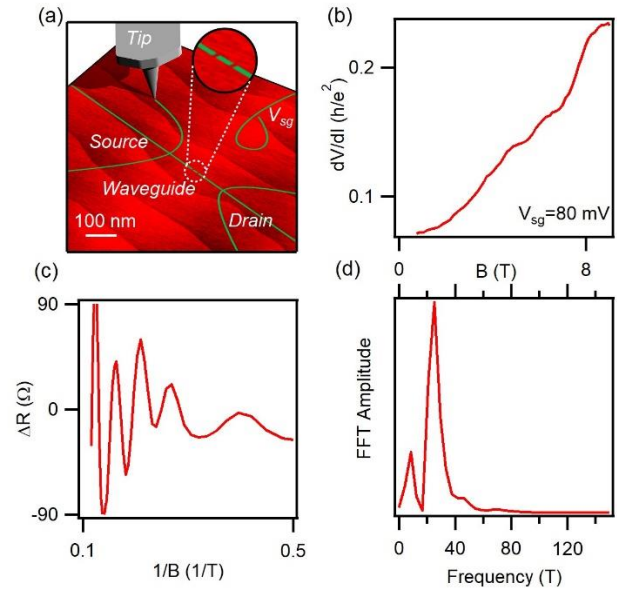

FIG. 1 The magnetoresistance of an electron waveguide. (a) Schematics of the electron waveguide. (b) The magnetoresistance at $V_{sg}$=80 mV. (c) SdH like oscillations after subtracting a smooth background from (b). (d) FFT analysis shows two distinct peaks.

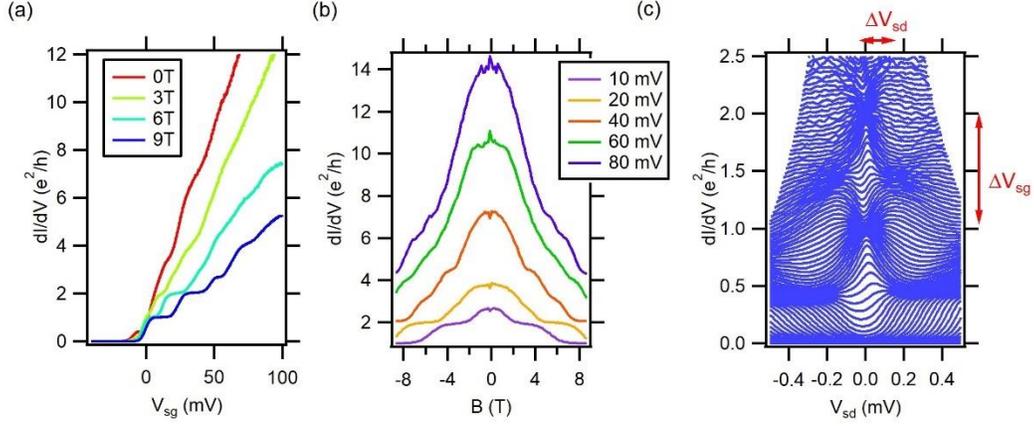

FIG. 2 Magnetic depopulation effect of the electron waveguide. (a) Differential conductance dependence on $V_{sg}$ in various magnetic fields $B$=0, 3, 6 and 9 T. Full conductance quantization is observed at high magnetic fields. (b) Magnetic depopulation for various $V_{sg}$ (10 mV to 80 mV). For a fixed $V_{sg}$, the number of occupied magnetoelectric bands decreases with increasing magnetic fields, signifying magnetic depopulation effect. (c) Finite-bias spectroscopy at 9 T. Differential *I-V* curves ranging from $V_{sg} = -10$ mV to 60 mV are clustering together where conductance is quantized. The arrows indicate the parameters ($\Delta V_{sd} \sim 0.22$ mV, $\Delta V_{sg} \sim 25.2$ mV) to extract the lever arm $\alpha = \frac{eV_{sd}}{V_{sg}} = 0.009$ eV/V.

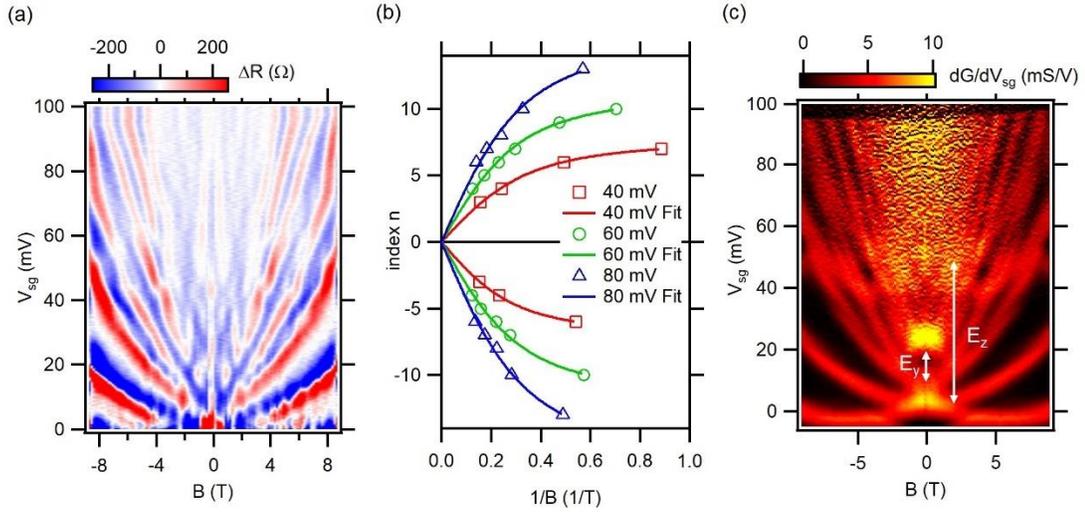

FIG. 3 Tunable magneto-electric subbands. (a) Resistance oscillations after subtracting a smooth background. (b) Subband index versus $1/B$ for $V_{sg}$=40 (red), 60 (green) and 80 mV (blue). Solid lines are fitting results. The subband indices are extracted by looking at the resistance minimum of (a). (c) Transconductance map showing lateral and vertical confinement energies $E_y \sim 100$ μeV (at $V_{sg} = 20$ mV) and $E_z \sim 500$ μeV (between the first two vertical bands), respectively.

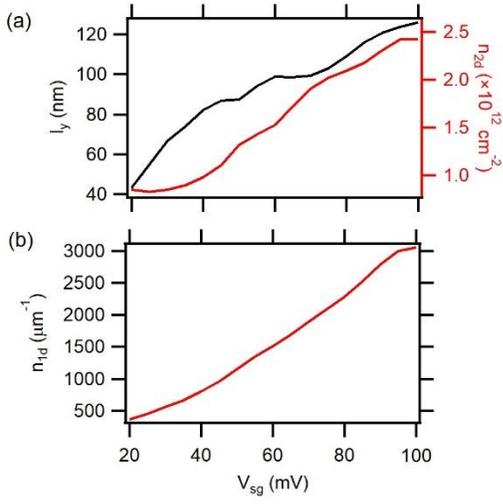

FIG. 4 Fitting results. (a) Extracted 2D carrier density and characteristic width of the waveguide. (b) Corresponding linear carrier density dependent on $V_{sg}$.